\documentclass[12pt]{revtex4}
\usepackage{eurosym}
\usepackage{amssymb}
\usepackage{epsfig}
\usepackage{graphicx}
\usepackage{amsmath}
\usepackage{amsfonts}
\usepackage{amssymb}
\usepackage{float}
\usepackage{subfigure}
\usepackage{float}
\usepackage{amsmath}
\usepackage{epstopdf}
\usepackage{color}
\usepackage{amsfonts}
\usepackage{amssymb}
\begin{document}

\title{\textbf{Holographic dark energy in modified Kaniadakis cosmology}}
\author{{\normalsize {\normalsize A. Sheykhi $^{1,2}$}\thanks{asheykhi@shirazu.ac.ir}, A. Asvar$^{1}$}\thanks{a\_asvar@pnu.ac.ir},
{\normalsize E. Ebrahimi$^{1,2}$}\thanks{es.ebrahimi@shirazu.ac.ir} \\
$^{1}${\normalsize Department of Physics, College of Science, Shiraz University, Shiraz 71454, Iran}\\
$^{2}${\normalsize Biruni Observatory, College of Science, Shiraz
University, Shiraz 71454, Iran}}

\begin{abstract}
It is well-known that any modification to the entropy expression
not only change the energy density of the holographic dark energy,
but also modifies the cosmological field equations through
thermodynamics-gravity correspondence. Here we propose the
Kaniadakis holographic dark energy (KHDE) in the background of the
modified Kaniadakis cosmology by incorporating the effects of
Kaniadakis entropy into the Friedmann equations. We choose the
Hubble radius, $L=H^{-1}$, as system's IR cutoff and determine the
cosmological implications of this model. We first consider a dark
energy (DE) dominated universe and reveal that this model mimics
the cosmological constant with $w_{DE}=-1$. This implies that the
theoretical origin of the cosmological constant, $\Lambda$, may be
understood through KHDE in the context of Kaniadakis cosmology.
Remarkably, we observe that in the absence of interaction between
DE and dark matter (DM), and in contrast to HDE in standard
cosmology, our model can explain the current acceleration of the
cosmic expansion for the Hubble radius as IR cutoff. When the
interaction between DE and DM is taken into account, we see that
the total equation of state parameter (EoS),
$w_{tot}=p_{tot}/\rho_{tot}$ can cross the phantom line at the
present time. We also analyze the squared speed of sound, $v_s^2$,
for this model and find out that $(v_s^2<0)$ for interacting KHDE.
Investigating the statefinder, confirms the distinction between
KHDE and $\Lambda$CDM model. It is seen that the statefinder
diagram move away from the point of $\left\lbrace
r,s\right\rbrace= \left\lbrace 1,0\right\rbrace$ with increasing
the interaction parameter.
\end{abstract}

\maketitle
\newpage
\newpage
\section{Introduction}\label{Int}
It is more than two decades that the foundation of the modern
cosmology has been shaken, after discovery of the acceleration of
the cosmic expansion by analyzing the data from high redshift type
Ia supernovaes \cite{riess, perlmutter}. In order to explain the
observed accelerated expansion, one needs to consider a new
component of energy with negative pressure $(p/\rho<-1/3)$. It is
well-known that the ordinary matter and radiation lead to a
decelerated universe and one should take into account a new
component of energy density usually called \textit{dark energy}.
The cosmological constant, $\Lambda$ with $w_{\Lambda}=-1$ is the
most acceptable DE candidate, because it is highly compatible with
observational evidences. However, cosmological constant suffers
some challenges such as fine-tuning and coincidence problems.
Recent observations imply discrepancies with the $\Lambda$CDM
which are known as cosmic tensions \cite{Lusso} and can be
summarized as $H_0$ \cite{Valentino} and $\sigma_8$
\cite{Riess2004,Riess2019} tensions. These challenges have
motivated the development of dynamical DE models as alternatives
to $\Lambda$CDM.

The holographic dark energy (HDE) model which is based on the
holographic principle admits time varying EoS parameter and has
received a lot of interests
\cite{Cohen,Hsu,Li,wang1,wang2,Wang2017}. The energy density of
HDE based on the holographic principle is defined as \cite{Cohen}
\begin{align}\label{rho1}
\rho_{DE}=3c^2M_p^2L^{-2}.
\end{align}
Here $3c^2$ is a numerical constant introduced for computational
convenience, $M_p^2=(8\pi G)^{-1}$ represents the reduced Planck
mass, and $L$ denotes the infrared (IR) cutoff, which is
associated with the scale of the observable universe. A natural
choice for $L$ is the Hubble radius, $L=1/H$. In a universe
dominated by a DE component, accelerated expansion occurs when the
EoS parameter satisfies $w_{DE}=p/\rho<-1/3$. However, it has been
noted that choosing $L=1/H$ and in the absence of interaction
between DE and DM, one finds $w_{DE}=0$ and thus the accelerated
expansion cannot be achieved \cite{Hsu,Li}. When the interaction
between DE and DM is taken into account, the simple choice for IR
cutoff could be the Hubble radius, $L=H^{-1}$ which can
simultaneously drive accelerated expansion and solve the
coincidence problem \cite{Pavon}. Many IR cutoffs have been
proposed in the context of HDE model. Choosing the future event
horizon as an IR cutoff, it was shown that the ratio of energy
densities can vary with time. With the interaction between the two
different constituents of the universe, the evolution of the
universe, from early deceleration to late time acceleration has
been explored in \cite{wang1,wang2}. The so called Granda-Olivers
(GO) cutoff is a combination of Hubble parameter and its first
derivative was proposed in HDE models \cite{Granda1,Granda2}. The
Ricci scalar is another IR cutoff which proposed for explanation
of the accelerated expansion \cite{Gao}.

On the other hand, the definition of energy density in the HDE
models crucially depends on the entropy associated with the
horizon. Any modification to the entropy expression will modify
the energy density of HDE. For example, Tsallis generalized
standard thermodynamics to non-extensive one, which can be applied
in all cases, and still possessing standard Boltzmann-Gibbs theory
as a limit \cite{Tsallis}. Based on this, and using the
statistical arguments, Tsallis and Cirto argued that the entropy
of a black hole does not obey the area law and can be modified as
\cite{Tsa} $S=\gamma A^{\beta}$ where $A$ is the horizon area,
$\gamma$ is an unknown constant, and $\beta$ known as
non-extensive parameter. The HDE models based on the modified
Tsallis entropy have been explored in \cite{Tav,Abd,Dut}. Another
modification to area law of entropy was proposed by Barrow
\cite{Barrow} who argued that the quantum gravitational effects
could cause changes in the geometry of the black hole horizon. The
change in the geometrical structure affects the form of the
horizon area and consequently the corresponding entropy expression
as well as the energy density of HDE. The HDE models based on the
modified Barrow entropy were addressed in
\cite{Saridakis2020,Srivastava,Adhikary:2021xym,Anag,Oliveros:2022biu}

The so called Kaniadakis entropy is a generalization of
Boltzmann-Gibbs-Shannon entropy that arises from relativistic
statistical theory and is characterized by a single parameter $K$
which quantifies the deviations from standard expressions
\cite{Kaniadakis2002,Kaniadakis2005}. This results from a
self-consistent and coherent relativistic statistical theory, in
which the basic features of standard statistical theory are
maintained. In such an extended statistical theory, the
distribution functions are a one-parameter continuous deformation
of the usual Maxwell-Boltzmann ones, and hence standard
statistical theory is recovered in a particular limit. HDE models
inspired by modified Kaniadakis entropy have been explored in the
literature \cite{Kum1,Dre,Kum2}. In particular, it was shown that
the EoS parameter of KHDE contains rich behavior, being
quintessence-like, phantom-like, or experience the phantom-divide
crossing in the past or in the future \cite{Dre}.

It is important to note that in most studies on KHDE, the authors
only modify the energy density of HDE, while they keep the
background Friedmann equations as in standard cosmology
\cite{Kum1,Dre,Kum2}. This is indeed incompatible with the
thermodynamics-gravity conjecture, which implies that any
modification to the entropy leads to modified dynamical field
equations of gravity. Therefore, in order to study the KHDE model,
one should take into account the modified energy density inspired
by Kaniadakis entropy, in the background of modified Kaniadakis
cosmology. The cosmological models through Kaniadakis corrected
entropy have been investigated in
\cite{Lym,Luciv,SheK1,SheK2,SheK3}. It was confirmed that the age
of the universe in the modified Kaniadakis cosmology becomes
larger compared to the standard cosmology which may alleviate the
age problem \cite{SheK2}. Besides, the transition from a
decelerated universe to an accelerated universe occurs at higher
redshifts compared to the standard cosmology \cite{SheK2}.

In the present work we shall revisit KHDE in the background of the
modified Kaniadakis cosmology.  Our work differs from
\cite{Kum1,Dre,Kum2} in that we consider KHDE with Hubble radius
as system's IR cutoff in the background of modified Kaniadakis
cosmology, while the authors of \cite{Kum1,Dre,Kum2} only modifies
the energy density of KHDE, and keep the Friedmann equations as in
standard cosmology.

The outline of this work is as follows. In the next section we
explore the cosmological model based on the modified Kaniadakis
entropy. We also study several cases of KHDE in the context of
Kaniadakis cosmology. In section III, we investigate, stability
and statefinder of the model. We finish with closing remarks in
the last section.
\section{KHDE in Modified Kaniadakis Cosmology}\label{Ther}
In this section we present the formalism of KHDE. The basic idea
behind HDE lies in the inequality ${\rho _{DE}}{L^4} \le S$ where
$L$ is the largest length scale of the theory or IR cutoff, and
$S$ represents the entropy of a black hole with radius $L$
\cite{Li,Wang2017}. For standard Bekenstein-Hawking entropy
${{\rm{\textit{S}}}_{{\rm{BH}}}} \propto {A
\mathord{\left/{\vphantom {A {(4G)}}} \right.
\kern-\nulldelimiterspace} {(4G)}} = {{\pi {L^2}}
\mathord{\left/{\vphantom {{\pi {L^2}} G}} \right.
\kern-\nulldelimiterspace} G}$. This implies that if instead of
standard Bekenstein-Hawking entropy, we use a modified one, we
obtain a modified energy density for HDE. As discussed in the
introduction, Kaniadakis entropy is a one-parameter generalization
of the classical entropy given by
\cite{Kaniadakis2002,Kaniadakis2005}
\begin{equation}
{S_K} =  - {k_B}\sum\limits_i {{n_{_i}}} {\ln _{\{ K\} }}{n_i},
\end{equation}
where ${k_B}$ is the Boltzmann constant and ${\ln _{\{ K\} }}x =
\left( {{x^K} - {x^{ - K}}} \right)/2K$. The Kaniadakis entropy is
characterized by the single dimensionless parameter $K$, which
quantifies deviation from standard statistical mechanics. The
standard entropy is recovered in the limit $K \to 0$ with $K$
constrained to $ - 1 < K < 1$. In this generalized statistical
theory the distribution function takes the form
\begin{equation}
{n_i} = \alpha \ {\exp _{\left\{ K \right\}}}\left[ { - \beta
\left( {{E_i} - \mu } \right)} \right],
\end{equation}
 where
 \begin{equation*}
{\exp _{\left\{ K \right\}}}\left( x \right) ={\left({\sqrt {1 +
{K^2}{x^2}}  + Kx} \right)^{{1 \mathord{\left/{\vphantom {1 K}}
\right. \kern-\nulldelimiterspace} K}}},
\end{equation*}
 \begin{equation*}
\alpha  = {\left[ {{{\left( {1 - K} \right)} \mathord{\left/
{\vphantom {{\left( {1 - K} \right)} {\left( {1 + K} \right)}}}
\right.\kern-\nulldelimiterspace} {\left( {1 + K} \right)}}}
\right]^{{1 \mathord{\left/{\vphantom {1 {2K}}} \right.
\kern-\nulldelimiterspace} {2K}}}},
\end{equation*}
and
\begin{equation*}
{1 \mathord{\left/{\vphantom {1 \beta }} \right.
\kern-\nulldelimiterspace} \beta } = \sqrt {1 - {K^2}} \  {k_\beta
}T.
\end{equation*}
Here the chemical potential $\mu$ can be fixed by normalization
\cite{Kaniadakis2002,Kaniadakis2005}. The Kaniadakis entropy can
also be expressed as \cite{Abreu2016,Abreu2017}
\begin{equation}\label{entropy1}
{S_K}= - {k_B}\sum\limits_{i = 1}^W {\frac{{P_i^{1 + K} - P_i^{1 -
K}}}{{2K}}},
\end{equation}
where ${P_i}$ is the probability of the system being in a specific
microstate and $W$ is the total number of configurations. In order
to apply Kaniadakis entropy in the black-hole framework, which is
essential for holographic application, we assume $P_i=1/W$. Using
the fact that Boltzmann-Gibbs entropy is $S \propto \ln W$, while
the Bekenstein-Hawking entropy is given by ${S_{BH}} = A/4G$, we
acquire $W = \exp \left( A/4G\right)$. Hereafter and for
simplicity, we work in a units system where ${k_B} = \hbar = c =
1$. Therefore, Eq.(\ref{entropy1}) can be written as
\begin{equation}\label{SK1}
{S_K} = \frac{1}{K}\sinh \left( {K{S_{BH}}} \right).
\end{equation}
When $K \to 0$ the standard Bekenstein-Hawking entropy is
restored, i.e. ${S_{K\to 0}} = {S_{BH}}$. The above modified
entropy expression should be close to the standard
Bekenstein-Hawking entropy, thus we assume $K \ll 1$. Thus, we can
expand the Kaniadakis entropy for small $K$, resulting
\begin{equation}\label{SK2}
{S_K} = {S_{BH}} + \frac{{{K^2}}}{6}S_{BH}^3 + {
\textit{O}}({K^4}).
\end{equation}
The first term is the usual Bekenstein-Hawking entropy, while the
second term is the lowest-order Kaniadakis correction term. The
energy density of KHDE, inspired by the modified entropy
(\ref{SK2}), is obtained as \cite{Dre}
\begin{equation}\label{energy density1}
{\rho _{DE}} = 3{c^2}M_p^2{L^{ - 2}} + 3 {\tilde c}^2 K^2
M_p^6{L^2},
\end{equation}
with $c$ and $\tilde c$ are arbitrary constants. For $K=0$ the
above expression gives the usual HDE density. According to the
above discussion, and using Eq. (\ref{energy density1}) with
$L=H^{-1}$, the energy density of KHDE can be rewritten as
\begin{equation}\label{energy density2}
{\rho _{DE}} = 3{c^2}M_p^2{H^2} + 3\alpha M_p^2H^{-2},
\end{equation}
where for later convenience we choose ${\tilde c}^2=32 \pi^4$ and
define
\begin{equation}
\alpha  \equiv \frac{{{K^2}{\pi ^2}}}{{2{G^2}}}.
\end{equation}
We assume the background spacetime is homogeneous and isotropic
with line elements
\begin{equation}
ds^2=-dt^2+a{(t)}^2\left[\frac{dr^2}{1-kr^2}+r^2(d\theta^2+\sin^2
(\theta)d\phi^2)\right],
\end{equation}
where $a(t)$ is the scale factor and $k$ denotes the curvature of
three-dimensional space. According to the thermodynamics-gravity
correspondence, the modified Friedmann equations inspired by
Kaniadakis entropy and in a flat background read
\cite{SheK1,SheK2}
\begin{eqnarray}\label{friedmann1}
&&3M_p^2\left( {{H^2} - \alpha {H^{ - 2}}} \right) = \left( {{\rho
_m} + {\rho _{DE}}} \right),\\
 &&- 2M_p^2\dot H\left( {1 + \alpha {H^{ - 4}}} \right) = \left( {{\rho _m} + {p_m} + {\rho _{DE}} + {p_{DE}}}
 \right),
 \end{eqnarray}
where $H=\dot{a}/a$, and ${{p_{DE}}}$ and ${{\rho_{DE}}}$ are
pressure and energy density of KHDE, respectively, while ${{\rho
_m}}$ and ${{p_m}}$ are, respectively, the energy density and
pressure of the matter sector. The total energy and pressure
satisfies the conservation equation
\begin{equation}
{\dot \rho} + 3H\left( {{\rho} + {p}} \right) = 0.
 \end{equation}
It is also convenient to introduce the density parameters as
usual,
\begin{eqnarray}\label{fractional matter density}
&&{\Omega _m} = \frac{{{\rho _m}}}{{{\rho _{cr}}}} = \frac{{{\rho
_m}}}{{3M_p^2{H^2}}}, \\
&&{\Omega _{DE}} = \frac{{{\rho _{DE}}}}{{{\rho _{cr}}}} =
\frac{{{\rho _{DE}}}}{{3M_p^2{H^2}}},\label{fractional energy
density}
\end{eqnarray}
where ${{\rho _{cr}} = 3M_p^2{H^2}}$ denotes the critical energy
density. Substituting the DE density from Eq. (\ref{energy
density2}) in Eq. (\ref{fractional energy density}), one finds
\begin{equation}\label{fractional energy density 2}
{\Omega _{DE}} = {c^2} + \alpha {H^{ - 4}}.
\end{equation}
\subsection{DE Dominated Universe}
As a special case, let us consider the late time universe where
the DE is dominated. Thus, we can neglect the contribution from DM
in the late time. Although, in reality one should consider two
dark sectors, however we leave it for the next subsections. For a
late time universe filled with $\rho_{DE}$, the first Friedmann
Eq. (\ref{friedmann1}) takes the following form
\begin{equation} \label{Fri1}
{H^2} - \alpha {H^{ - 2}} = \frac{{{\rho _{DE}}}}{{3M_p^2}}.
\end{equation}
Substituting relation (\ref{energy density2}) into Eq.
(\ref{Fri1}), we have
\begin{equation}
{H^2}(1 - {c^2}) = 2\alpha {H^{ - 2}},
\end{equation}
which admits the following solution for the Hubble parameter
\begin{equation} \label{H}
H = \left[\frac{2\alpha}{1-c^2} \right]^{1/4}=\rm constant.
\end{equation}
Thus the constant $c^2$  must be lower than unity ($c^2<1$), which
is consistent with previous studies \cite{Li,Pavon}. The
corresponding scale factor becomes
\begin{equation}
a(t)\sim \exp \left[\left(\frac{2\alpha }{1 - c^2} \right)^{1/4} t
\right].
\end{equation}
Using the continuity equation for KHDE, ${\dot \rho _{DE}} +
3H{\rho _{DE}}\left( {1 + w_{DE}} \right)=0$, we find the EoS
parameter for KHDE as
\begin{equation}
w_{DE} = - 1.
\end{equation}
Thus in this case the KHDE mimics the cosmological constant, our
universe undergoes an accelerated expansion phase, and evolves
like a pure de-Sitter space. This is an interesting result and
shows that the correction term comes from Kaniadakis entropy into
the dynamical field equations, can bring rich physics compared to
the standard cosmology and even compared to the previous studies
in KHDE models \cite{Kum1,Dre,Kum2}. Note that in the absence of
correction term, $\alpha=0=K^2$, we have a static universe with
$H=0$.

Another interesting result in this case comes from energy density
of KHDE. Substituting $H$ from (\ref{H}) into (\ref{energy
density2}) we see that $\rho_{DE}=\rm constant, $ as expected for
a pure de Sitter spacetime. Equating the energy density of KHDE
with energy density of cosmological constant $\Lambda$, namely
$\rho_{DE}=\rho_{\Lambda}=\Lambda/(8 \pi G)$ for the case of DE
dominated universe, one can easily find
\begin{equation}
\Lambda=\frac{3 \pi K}{2G}\frac{1+c^2}{\sqrt{1-c^2}}.
\end{equation}
As a result $\Lambda\sim K$, which confirms that $K$ is very
small. This may provide a theoretical origin for the cosmological
constant through thermodynamic arguments. In other words, the
origin of $\Lambda$ can be understood from Kaniadakis modified
entropy. This is one of the novel result of the present work.
\subsection{Noninteracting Case}
In the previous subsection, we explored KHDE model for DE
dominated epoch and neglected the contribution from matter sector.
Here, we would like to consider the KHDE model in an epoch that
the cosmic fluid includes DM  and DE components simultaneously. At
first, we assume that these components are separately conserved.
The conservation equations when matter and DE evolve separately
read
\begin{eqnarray}\label{energy conservation}
&&{\dot \rho _{DE}} + 3H{\rho _{DE}}\left( {1 + {w_{DE}}} \right)
=0,\\ &&{\dot \rho _m} + 3H{\rho _m} = 0. \label{matter
conservation}
\end{eqnarray}
We can redefine the DE density as
$\rho_{DE}^{new}=3{c^2}M_p^2{H^2} + 3\alpha^{\prime} M_p^2H^{-2}$,
with $\alpha^{\prime}=2\alpha$. Thus the first Friedmann equation
reads $3M_p^2{H^2}=\rho _m+\rho_{DE}^{new}$. Since $\alpha$ and
$\alpha^{\prime}$ are arbitrary constants, hence without loss of
generality we can omit the prime and write
$\rho_{DE}^{new}=\rho_{DE}$. Therefore we can rewrite the first
Fridmann equation as
\begin{equation}\label{om}
{\Omega _m} + {\Omega _{DE}} = 1.
\end{equation}

Taking the time derivative of the first Friedmann equation and
substituting ${\dot \rho _m}$ and ${\dot \rho _{DE}}$ from the
conservation Eqs. (\ref{energy conservation}) and (\ref{matter
conservation}), we arrive at
\begin{equation}
\frac{{\dot H}}{{{H^2}}} =  - \frac{3}{2}\left( {1 +
{w_{DE}}{\Omega _{DE}}} \right).
\end{equation}
Inserting  ${\dot \rho _{DE}}$ into Eq. (\ref{energy
conservation}) after some algebra, we find the EoS parameter as
\begin{equation}\label{wd1}
w_{DE} = \left( {\frac{2}{{{\Omega _{DE}}}}} \right)\left(
{\frac{{{c^2} - {\Omega _{DE}}}}{{1 + {\Omega _{DE}} - 2{c^2}}}}
\right).
\end{equation}
In the limiting case where $\alpha=0=K^2$, from Eq.
(\ref{fractional energy density 2}), we find $\Omega_{DE}=c^2$ and
hence $w_{DE}=0$. This implies that in the context of standard
cosmology and the absence of interaction between DE and DM,
choosing $L=H^{-1}$ does not lead to an accelerated expansion
universe \cite{Hsu}. However, from Eq. (\ref{wd1}) we see that in
the context of Kaniadakis cosmology, choosing Hubble radius as IR
cutoff can lead to an accelerated universe. This is an interesting
and novel result and shows that the entropy corrected KHDE
combined with correct dynamical background equations bring rich
physics. Requiring the fact that $w_{DE}<-1/3$, from Eq.
(\ref{wd1}) we find that constant $c^2$ should be bounded as
\begin{equation}\label{cc}
c^2<\frac{\Omega _{DE}}{2}\left(\frac{5-\Omega _{DE}}{3-\Omega
_{DE}}\right).
\end{equation}
Thus at the present time where $\Omega_{DE}\simeq 0.7$, we find
$c^2\lesssim 0.654$. Next we calculate the total EoS parameter,
which is defined as
\begin{equation}\label{wtot1}
{w_{tot}} = \frac{{{p_{DE}}}}{{{\rho _m} + {\rho _{DE}}}}.
\end{equation}
Using ${\rho _m} = {\rho _{m0}}{a^{ - 3}}$ and ${w_{DE}} =
{{{p_{DE}}} \mathord{\left/
        {\vphantom {{{p_{DE}}} {{\rho _{DE}}}}} \right.
        \kern-\nulldelimiterspace} {{\rho _{DE}}}}$, we get

\begin{equation}\label{wtot2}
{w_{tot}} = \frac{{{w_{DE}}}}{{{{{\rho _m}_0{a^{ - 3}}} \mathord{\left/
                {\vphantom {{{\rho _m}_0{a^{ - 3}}} {{\rho _{DE}}}}} \right.
                \kern-\nulldelimiterspace} {{\rho _{DE}}}} + 1}}.
\end{equation}
Given ${{\rho _{DE}}}$ from Eq. (\ref{energy density2}) and
${{w_{DE}}}$  from Eq. (\ref{wd1}), one can calculate $w_{tot}$
provided the value of $H$ and the evolutionary form of $\Omega
_{DE}$ are known. Substituting Eq. (\ref{energy density2}) into
Eq. (\ref{friedmann1}), we arrive at the following equation
\begin{equation}\label{quadratic equation}
3M_p^2\left( {1 - {c^2}} \right){H^4} - {\rho _m}{H^2} - 3\alpha
M_p^2 = 0,
\end{equation}
which admits the following solution for the Hubble parameter
\begin{equation}\label{second power Hubble parameter}
{H^2} = \frac{{{\rho _{m0}}{a^{ - 3}} \pm \sqrt {{\rho
_{m0}}^2{a^{ - 6}}+ 36\alpha M_p^4\left( {1 - {c^2}} \right)}
}}{{6M_p^2\left( {1 - {c^2}} \right)}}.
\end{equation}
Since the radical term is larger than the first term in the
fraction, only the positive sign is acceptable. Replacing Eq.
(\ref{second power Hubble parameter}) in Eq. (\ref{energy
density2}) and then in Eq. (\ref{wtot2}), we arrive at

\begin{equation}\label{Total state equation parameter4}
{w_{tot}} = {w_{DE}}\left[ {\frac{{3{c^2}M_p^2\left( {{{{H^2}} \mathord{\left/
                        {\vphantom {{{H^2}} {{\rho _{cr0}}}}} \right.
                        \kern-\nulldelimiterspace} {{\rho _{cr0}}}}} \right) + 3\alpha M_p^2\left( {{{{H^{ - 2}}} \mathord{\left/
                        {\vphantom {{{H^{ - 2}}} {{\rho _{cr0}}}}} \right.
                        \kern-\nulldelimiterspace} {{\rho _{cr0}}}}} \right)}}{{{\Omega _{m0}}{{\left( {1 + z} \right)}^3} + 3{c^2}M_p^2\left( {{{{H^2}} \mathord{\left/
                        {\vphantom {{{H^2}} {{\rho _{cr0}}}}} \right.
                        \kern-\nulldelimiterspace} {{\rho _{cr0}}}}} \right) + 3\alpha M_p^2\left( {{{{H^{ - 2}}} \mathord{\left/
                        {\vphantom {{{H^{ - 2}}} {{\rho _{cr0}}}}} \right.
                        \kern-\nulldelimiterspace} {{\rho _{cr0}}}}} \right)}}}
                        \right],
\end{equation}
where  ${\rho _{cr0}} = 3M_p^2H_0^2$, $a = {\left( {1 + z}
\right)^{ - 1}}$ with $z$ being the redshift parameter. Here
${H_0}$ and ${\Omega _{m0}} = {{{\rho _{m0}}} \mathord{\left/
{\vphantom {{{\rho _{m0}}} {{\rho _{cr0}}}}} \right.
\kern-\nulldelimiterspace} {{\rho _{cr0}}}}$ are the values of the
Hubble parameter and fractional matter density at the present
time, respectively. Besides, we can write terms $\left( {{{{H^2}}
\mathord{\left/{\vphantom {{{H^2}} {{\rho _{cr0}}}}} \right.
\kern-\nulldelimiterspace} {{\rho _{cr0}}}}} \right)$ and $\left(
{{{{H^{ - 2}}} \mathord{\left/{\vphantom {{{H^{ - 2}}} {{\rho
_{cr0}}}}} \right. \kern-\nulldelimiterspace} {{\rho _{cr0}}}}}
\right)$ as
\begin{equation}\label{D1}
\frac{{{H^2}}}{{{\rho _{cr0}}}}= \frac{{{\Omega _{m0}}{{\left( {1
+ z} \right)}^3} + \sqrt {\Omega _{m0}^2{{\left( {1 + z}
\right)}^6} + {{36\alpha M_p^4\left( {1 - {c^2}} \right)}
\mathord{\left/   {\vphantom {{36\alpha M_p^4\left( {1 - {c^2}}
\right)} {\rho _{cr0}^2}}} \right.\kern-\nulldelimiterspace} {\rho
_{cr0}^2}}} }}{{6M_p^2\left( {1 - {c^2}} \right)}},
\end{equation}
\begin{equation}\label{D2}
\frac{{{H^{ - 2}}}}{{{\rho _{cr0}}}} = \frac{{6M_p^2\left({1 -
{c^2}} \right)}}{{{\Omega _{m0}}{{\left( {1 + z} \right)}^3}\rho
_{cr0}^2 + \sqrt {\Omega _{m0}^2{{\left( {1 + z} \right)}^6}\rho
_{cr0}^4 + 36\alpha M_p^4\left( {1 - {c^2}} \right)\rho _{cr0}^2}
}}.
\end{equation}
In what follows, we use $x = \ln a$ as an independent variable and
therefore we denote derivation with respect to $x$ by $\prime$
(and also note that $\dot f = f'H$). Differentiating Eq.
(\ref{fractional energy density 2}) in terms of \textit{x} and
using Eqs. (\ref{energy conservation}) and (\ref{matter
conservation}), we obtain
\begin{equation}\label{fractional energy density 2 prime}
{\Omega '_{DE}} = 6\left( {{\Omega _{DE}} - {c^2}} \right)\left( {\frac{{1 - {\Omega _{DE}}}}{{1 + {\Omega _{DE}} - 2{c^2}}}} \right).
\end{equation}
In order to obtain derivative of ${{\Omega _{DE}}}$ with respect to redshift parameter $z$, one should note
\begin{equation}\label{derivative of fractional energy density 2 prime}
\frac{{d{\Omega _{DE}}}}{{dz}} = \frac{{d{\Omega _{DE}}}}{{dx}}\frac{{da}}{{adz}} =-\frac{1}{{1 + z}}{\Omega'_{DE}}.
\end{equation}
Substituting Eq. (\ref{fractional energy density 2 prime}) into
Eq. (\ref{derivative of fractional energy density 2 prime}), one
gets
\begin{equation}
\frac{{d{\Omega _{DE}}}}{{dz}} = \frac{{6({c^2}-{\Omega _{DE}})}}{{1 + z}}\left[ {\frac{{\left( {1 - {\Omega _{DE}}} \right)}}{{1 + {\Omega _{DE}} - 2{c^2}}}} \right].
\end{equation}
The deceleration parameter is obtained as
\begin{equation}
q =  - 1 - \frac{{\dot H}}{{{H^2}}} =  - 1 + \frac{3}{2}\left(
{\frac{{1 - {\Omega _{DE}}}}{{1 + {\Omega _{DE}} - 2{c^2}}}}
\right).
\end{equation}
In the limiting case where $\alpha$ is equal to zero ($K=0$), the
parameters $\Omega_{DE}$, $w_{DE}$, $\Omega'_{DE}$, and $q$ are
respectively equal to $c^2$, $0$, $0$, $1/2$, revealing that the
standard HDE with Hubble radius as IR cutoff, does not lead to an
accelerated universe, as we expect.

It is important to note that the Kaniadakis entropy is an even
function, namely ${S_K}={S_{ - K}}$, and that is why all the above
expressions for KHDE depend only on ${K^2}$. The details
discussion on the behaviour of the cosmological parameters for
noninteracting KHDE will be given in the following subsection,
where we plot the figures for both interacting and noninteracting
cases.
\subsection{Interacting Case}
Next, we analyze a flat universe filled with KHDE and DM, while
swapping energy among them. Since the microscopic nature of both
DE and DM is still unknown, there exist enough room to leave a
chance of interaction between the two dark sectors. Besides, it is
shown that in the presence of interaction between dark components,
the coincidence problem could be alleviated \cite{Pavon}. In
\cite{SheyCQ}, the interaction between DM and DE has been
investigated from a thermodynamic point of view. Thus, there exist
enough motivations and interests to explore an interacting version
of dark components. The conservation equations including an
interaction between DM and DE, read
\begin{eqnarray}\label{cons1}
&&{\dot \rho _{DE}} + 3H{\rho _{DE}}\left( {1 + w_{DE}} \right) =
-Q, \\ && {\dot \rho _m} + 3H{\rho _m} = Q,\label{cons2}
\end{eqnarray}
where $Q=3b^2H(1+r)\rho_{DE}$ is the interaction term
\cite{Sheykhi2011}. In this relation $b^2$ is a coupling constant
and $r=\rho_{m}/\rho_{DE}$ \cite{Pavon}. Negative sign of $Q$
indicates a transfer of energy from DE to DM. Following the same
steps as the previous subsection, one can find the resulting
equations for $w_{DE}$, $q$ as well as the evolution equation of
$\Omega_{DE}$. Taking derivative of Friedmann equations and
substituting ${\dot \rho _m}$ and ${\dot \rho _{DE}}$ from the
conservation Eqs. (\ref{cons1}) and (\ref{cons2}), we find
\begin{equation}
\frac{{\dot H}}{{{H^2}}} =-\frac{3}{2}\left({1 + {w_{DE}}{\Omega
_{DE}}} \right).
\end{equation}
In this case, it is a matter of calculations to show that
\begin{equation}\label{wd2}
{w_{DE}} = \frac{2}{\Omega _{DE}}\left(\frac{c^2- \Omega _{DE}-
b^2/6}{1+\Omega _{DE} - 2c^2}\right).
\end{equation}
The total EoS is still obeyed from Eq. (\ref{wtot2}) where
$w_{DE}$ is now given by (\ref{wd2}). Differentiating Eq.
(\ref{fractional energy density 2}) in terms of \textit{x} and
using Eqs. (\ref{cons1}) and (\ref{cons2}), we obtain
\begin{equation}
{\Omega '_{DE}} = 3\left( {{\Omega _{DE}} - {c^2}}\right)\left[
{\frac{{3(1 - {\Omega _{DE}}) - {b^2}}}{{1 + {\Omega _{DE}} -
2{c^2}}}} \right].
\end{equation}
Then we obtain evolution equation of the HDE in Kaniadakis
cosmology
\begin{equation}
\frac{{d{\Omega _{DE}}}}{{dz}}= \frac{{3\left({{c^2}-{\Omega
_{DE}}} \right)}}{{1 + z}}\left[ {\frac{{3\left( {1 - {\Omega
_{DE}}} \right) - {b^2}}}{{1 + {\Omega _{DE}} - 2{c^2}}}} \right].
\end{equation}
Also, the deceleration parameter can be obtained as
\begin{equation}
q =-1+\frac{3}{2}\left(\frac{1 -\Omega _{DE}-b^2/3}{1+\Omega
_{DE}-2c^2}\right).
\end{equation}
When $b^2=0$, the results of the previous subsection are restored.
In the limiting case where $\alpha=0=K^2$, we have
$\Omega_{DE}=c^2$, $\Omega'_{DE}=0$, and
\begin{equation}
w_{DE}=-\frac{b^2}{3c^2(1-c^2)}, \   \   \   \    \
q=\frac{1}{2}\left(1-\frac{b^2}{1-c^2}\right).
\end{equation}
This implies that in the context of standard cosmology, any
interaction of pressureless DM with HDE, whose infrared cutoff is
set by the Hubble scale, can explain the accelerated expansion and
solve the coincidence problem \cite{Pavon}. When $\alpha=0=b^2$,
we have $w_{DE}=0$, implying that in standard cosmology with
$L=H^{-1}$, interaction is the only way to reproduce an
accelerated universe. As we discussed in the previous subsection,
in the context of Kaniadakis cosmology, however, the accelerated
expansion can be achieved for KHDE, even in the absence of
interaction between DM and DE.

\begin{figure}[htp] \label{fig1}
\centering
\begin{tabular}{cc}
 \includegraphics[width=75mm]{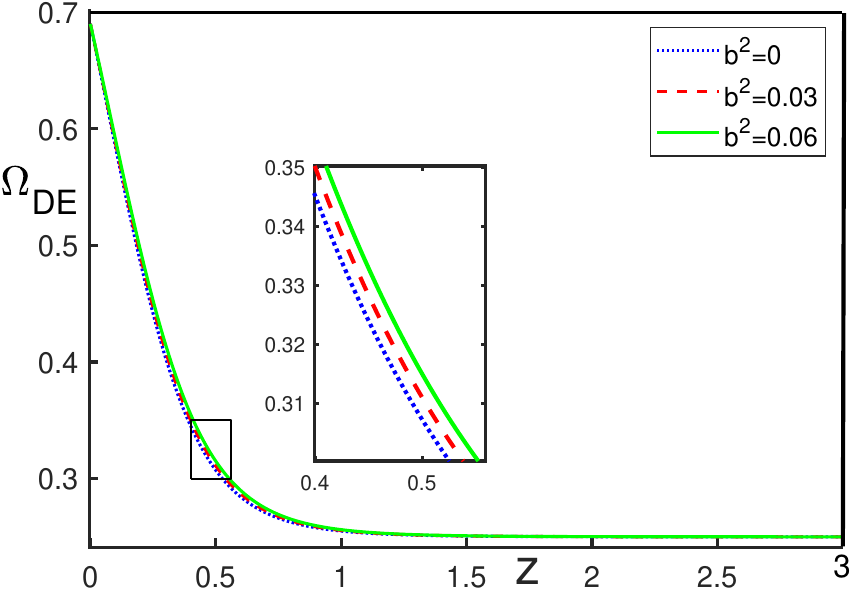}
 \includegraphics[width=75mm]{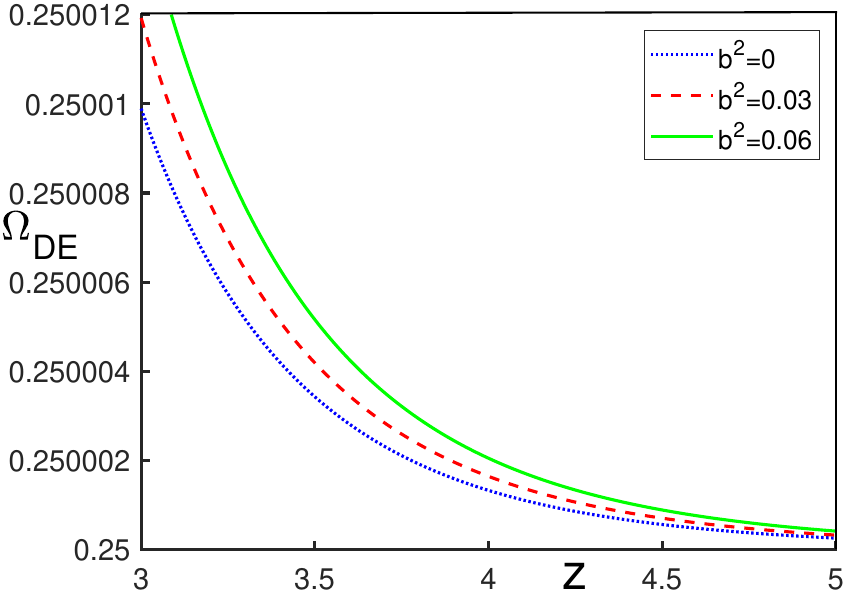}
 \end{tabular}
\caption{\scriptsize Evolution of $\Omega_{DE}$ (left panel) and
(right panel) as a function of redshift, $z$, for both interacting
and noninteracting KHDE for different  values of ${b^2}$ in a flat
Kaniadakis cosmology. Here, we set $\Omega^{0}_{DE}$= 0.7, and
${c^2} = 0.5$.}
\end{figure}

We plot the evolution of $\Omega_{DE}$ with respect to redshift in
Fig.\ref{fig1} for both interacting and noninteracting KHDE. From
this figure (left panel), we observe that for $0<z<1.5$, the value
of $\Omega_{DE}$ decreases significantly with the increase of
redshift while for $(z>1.5)$ it is almost constant. The evolution
of $\Omega_{DE}$ versus redshift in the range of $3<z<5$ for both
interacting and noninteracting cases are plotted in
Fig.\ref{fig1}(right panel). The right part of Fig.\ref{fig1}
complements the behavior of $\Omega_{DE}$ at large values of $z$.

\begin{figure}[htp]\label{fig2}
 \centering
 \begin{tabular}{cc}
 \includegraphics[width=120mm]{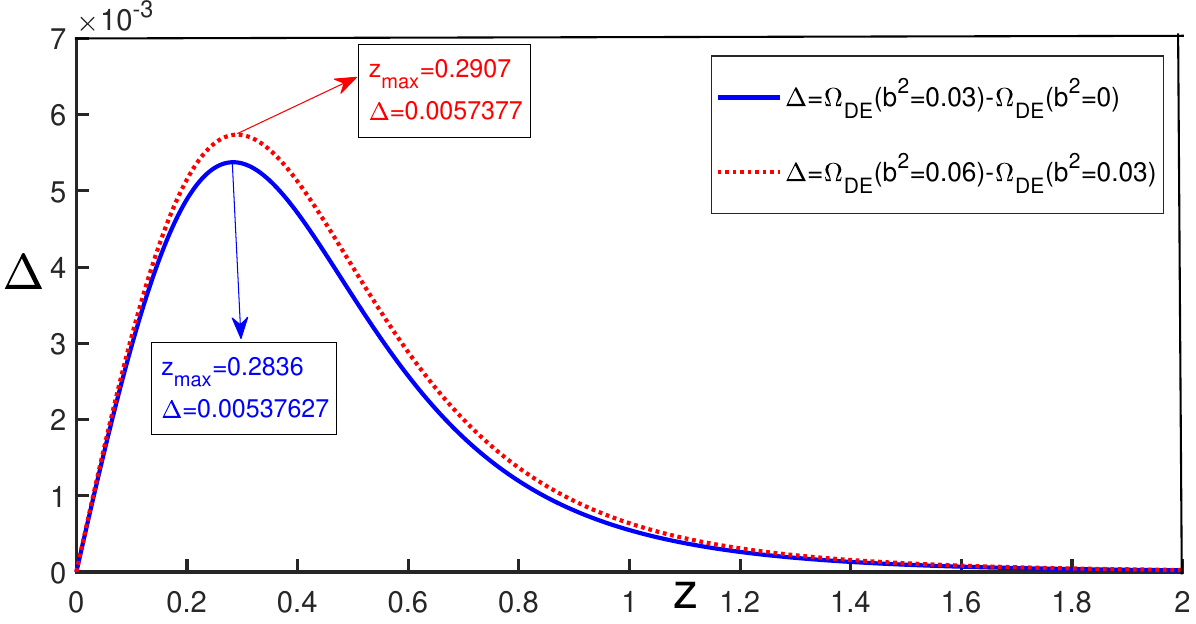}
\end{tabular}
\caption{\scriptsize Evolution of the difference between the
$\Omega_{DE}$ values ($\Delta$) as a function of $z$, for
different vales of ${b^2}$ in a flat Kaniadakis cosmology. Here,
we set $\Omega^{0}_{DE}$= 0.7, and  ${c^2} = 0.5$. }
\end{figure}

We also for more clarity obtain the difference between the
$\Omega_{DE}$ values,($\Delta$), for different values of
interacting parameter. It is evident from Fig.\ref{fig2} that as
$z$ increases from zero, the difference between $\Omega_{DE}$
curves grows, reaching a maximum at $z_{max}$. Finally, for $z$
values greater than $z_{max}$, the difference function decreases
with increasing $z$ and asymptotically approaches zero.

The evolutions of $w_{DE}$ (left panel) and $w_{tot}$ (right
panel) with respect to redshift $z$ for KHDE with Hubble horizon
IR cutoff are plotted in Fig.\ref{fig3}. Let us probe in more
detail the EoS parameter in Fig. \ref{fig3} (left panel). An
overall harvest is that with the increase of redshift for
$0<z<1.5$, the EoS parameter $w_{DE}$ increases while for
$(z>1.5)$ seems constant. According to this figure, $w_{DE}$
parameter at present time crosses the phantom divide, namely,
$w_{DE} < - 1$  in the absence of interaction. Moreover, from the
left graph of Fig.(\ref{fig3}) we can see that the value of
$w_{DE}$ at present epoch is around ($-1$) which is favored by
observations. We also see that for $0<z<0.6$, the total EoS
parameter increases with increasing $z$, while for $(z>0.6)$ it
becomes constant. We can observe that $w_{tot}$ of the KHDE
remains in quintessence era, and approaches to $w_{tot}=-1$ at
future for non-interacting case implying a big rip singularity as
fate of the universe.

Let us now look at Fig. 3 for interacting case. A close look to
the left panel of Fig.\ref{fig3} reveals that with increasing the
coupling constant $b^2$, the EoS parameter of KHDE decreases as
well. Besides, for all choices of $b^2$, $w_{DE}$ lies in the
phantom regime at the present epoch. Moreover, from the left graph
of Fig.\ref{fig3} we can that at the present time,
$-1.2<w_{DE}<-1$ which is in agreement with observations
\cite{Planck2018}. We also see that at a given $z$, $w_{tot}$,
decreases with increasing $b^2$. While $w_{tot}$ remains in
quintessence era in the non interacting case ($b^2=0$) and its
value approaches to $w_{tot}=-1$ at future but for $b^2= 0.03 ,
0.06$ it mildly crosses the border line $w_{tot}=-1$.

In summary, we can see that the scenario of KHDE can lead to very
interesting cosmological phenomenology, in which $w_{DE}$ and
$w_{tot}$ can be quintessence-like, phantom-like, or cross the
phantom divide during the history of the universe. Note that these
results are in agreement with other researches on KHDE, that
appeared after the present work, especially with observational
confrontation \cite{Hern1,Hern2,Lymperis,Luciano,Ghaffari}.

 \begin{figure}[htp] \label{fig3}
    \centering
    \begin{tabular}{cc}

 \includegraphics[width=75mm]{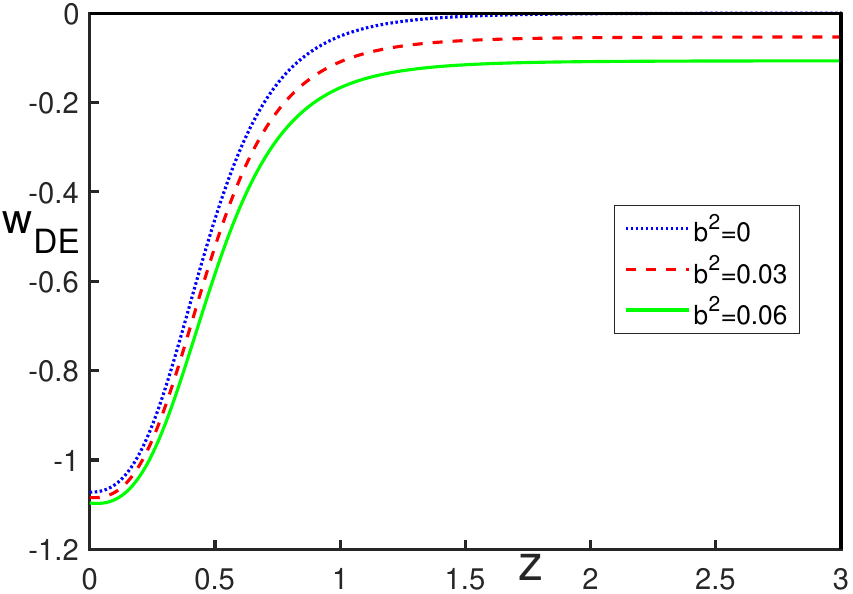}

 \includegraphics[width=75mm]{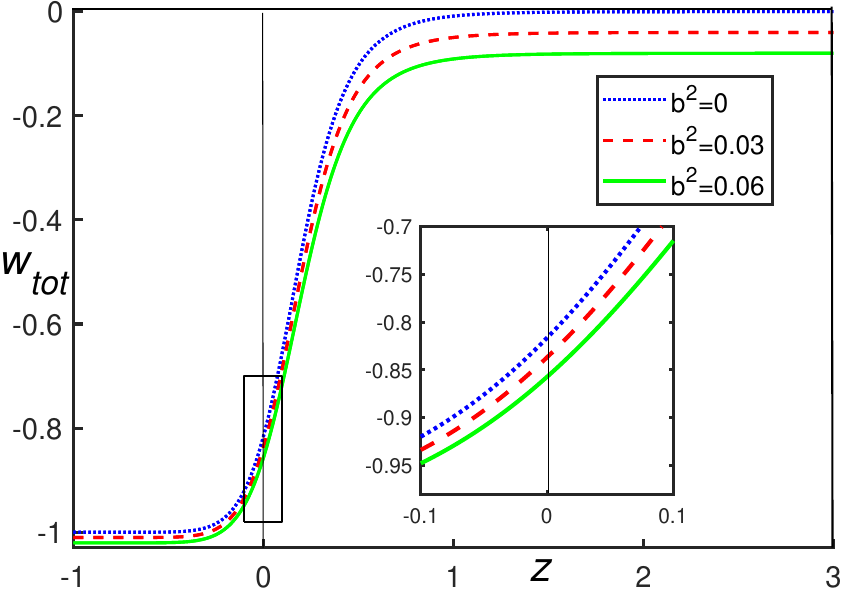}

    \end{tabular}
\caption{\scriptsize Evolution of $w_{DE}$ (left panel) and
$w_{tot}$ (right panel) as a function of $z$ for different choices
of ${b^2}$ in a flat Kaniadakis cosmology. Here, we set $K=0.1$,
$\Omega_{DE0}$= 0.7, and ${c^2} = 0.5$. }\label{fig3}
 \end{figure}
\begin{figure}[htp]\label{fig4}
\centering
\begin{tabular}{cc}
\includegraphics[width=90mm]{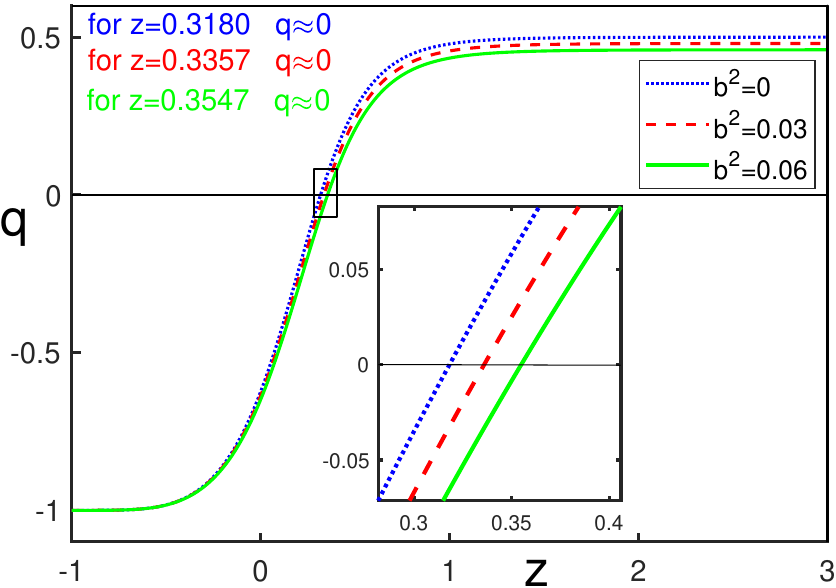}
\end{tabular}
\caption{\scriptsize Evolution of the deceleration parameter $q$
as a function of redshift, $z$, for KHDE in a flat Kaniadakis
cosmology. Here, we set $\Omega_{DE0}$= 0.7, and  ${c^2} = 0.5$. }
\end{figure}

The evolution of the deceleration parameter with respect to
redshift, $z$, is presented in Fig.\ref{fig4}. Note that $\left(
{q < 0} \right)$, implies an accelerating universe while $\left(
{q > 0} \right)$, denotes a decelerating phase. It is proposed by
observations that the universe is in an accelerated expansion
phase and the value of deceleration parameter lies in the range $
- 0.6 \le q < 0$ \cite{Planck2018}. We observe from this figure
that the deceleration parameter of the KHDE model transits from an
early decelerated phase to the current accelerated phase for
noninteracting case ($b^2=0$). The transition from a decelerated
universe to the accelerated expansion happens around $z\approx
0.4$. This result indicates a delay in phase transition in
comparison with respect to $\Lambda$CDM where predicts
$z_{tr}=0.64\pm 0.04$ \cite{Planck2018}. When the interaction
between DE and DM is taken into account, the situation becomes
better and the phase transition occurs at larger $z$. According to
the curves, when the interaction between DM and DE increases, the
deceleration parameter decreases at an special redshift.
\section{Stability and Statefinder of the Model}\label{Ther}
\subsection{Stability}
Squared sound speed $\upsilon _s^2$ analysis, is an interesting
method which reveals signs of gravitational instability through a
semi-Newtonian approach. In an expanding background, deep inside
the horizon, one can find evolution of the cosmic perturbations as
\begin{equation}
\delta \rho  \propto {e^{ \pm i\omega t}},
\end{equation}
where $\omega  \propto {\upsilon _s}$ and $\delta \rho$ denotes
the density perturbations. The negative and positive signs of
$\upsilon _s^2$ could be a sign of instability and stability
respectively. To proceed the analysis, we start with definition of
the adiabatic squared sound speed which reads
\begin{equation}\label{squared sound speed}
\upsilon _s^2 = \frac{{d{p_{DE}}}}{{d{\rho _{DE}}}}
\end{equation}
where ${p_{DE}}$ and ${\rho _{DE}}$ are the pressure and the
density of the DE component, respectively. According to ${p_{DE}}
= {\rho _{DE}}{w_{DE}}$, Eq.(\ref{squared sound speed}) can be
written as
\begin{equation}
\upsilon _s^2 = \frac{\dot{\rho _{DE} w_{DE}}}{{{{\dot \rho
}_{DE}}}} = {w_{DE}}+ \frac{{{\rho _{DE}}}}{{{{\dot \rho
}_{DE}}}}\frac{{d{w_{DE}}}}{{d{\Omega _{DE}}}}\frac{{d{\Omega
_{DE}}}}{{dt}}.
\end{equation}
Substituting ${{{\rho _{DE}}} \mathord{\left/ {\vphantom {{{\rho
_{DE}}} {{{\dot \rho }_{DE}}}}}\right.\kern-\nulldelimiterspace}
{{{\dot \rho }_{DE}}}}$ from the conservation Eq.(\ref{cons1}),
one may get
\begin{equation}
\upsilon _s^2 = {w_{DE}}+ \frac{{d{w_{DE}}}}{{d{\Omega
_{DE}}}}\left[ {\frac{{\left( {{c^2} - {\Omega _{DE}}}
\right)\left( {3\left( {1 - {\Omega _{DE}}} \right) - {b^2}}
\right)}}{{\left( {1 + {w_{DE}} + {{{b^2}} \mathord{\left/
{\vphantom {{{b^2}} {{\Omega _{DE}}}}} \right.
\kern-\nulldelimiterspace} {{\Omega _{DE}}}}} \right)\left( {1 +
{\Omega _{DE}} - 2{c^2}} \right)}}} \right].
\end{equation}
We plot the evolution of $\upsilon _s^2$ as a function of $z$ for
different values of ${b^2}$ in a flat background in
Fig.\ref{fig5}. It can be seen that with increasing of ${b^2}$ the
stability of this model decreases. It is interesting to note that
$\upsilon _s^2$ can takes positive values in the past while the
model shows stable phase at present epoch for $b^2=0.06$. However
in this model, evidences of instability is seen in the
presence(absence) of an interaction between DM and DE during
evolution of the Universe. An overall result is that the KHDE
suffers the stability problem.
\begin{figure}[htp]\label{fig5}
    \centering
 \begin{tabular}{cc}
 \includegraphics[width=80mm]{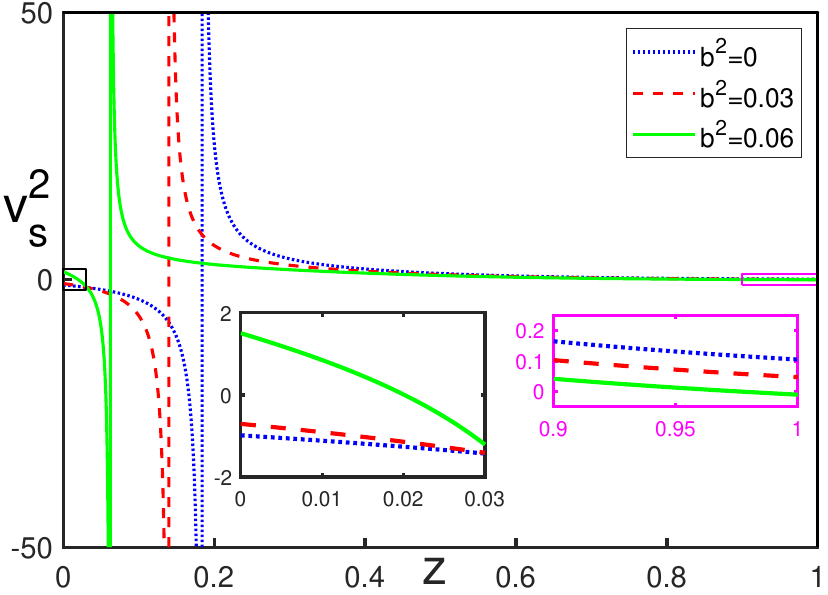}

    \end{tabular}
\caption{\scriptsize Evolution of the squared sound speed
$\upsilon _s^2$ as a function of redshift, $z$, for
non-interacting and interacting KHDE in a flat Kaniadakis
cosmology. Here, we set $\Omega_{DE0}$= 0.7, and ${c^2} = 0.5$. }
\label{fig5}
\end{figure}
\subsection{Statefinder}
Sahni et al. introduced the statefinder pair $\left\{ {r,s}
\right\}$, which includes third time derivative of the scale
factor $a\left( t \right)$ \cite{Sahni}. Since DE models are very
diverse, it is useful and essential to distinguish between
different DE models. The statefinder pair $\left\{ {r,s} \right\}$
is an effective tool for finding differences between DE models and
the standard model \textit{$\left( {\Lambda CDM} \right)$}. It
should be noted, for the standard model \textit{$\left( {\Lambda
CDM} \right)$} the statefinder parameters are $\left\{ {r,s}
\right\} = \left\{ {1,0} \right\}$. The statefinder pair $\left\{
{r,s} \right\}$ is defined as \cite{Colg,Hossien}
\begin{eqnarray}
&&\label{rr}
r=\dfrac{\dddot{a}}{aH^3}=2q^2+q-\dfrac{\dot{q}}{H},\\
&&\label{s} s=\dfrac{r-1}{3[q-(1/2)]}.
\end{eqnarray}
Based on the relationship between $r$ and the deceleration
parameter $q$, the $r$ parameter could be obtained as
\begin{equation}
r= 2{q^2} + q - \frac{1}{H}\frac{{dq}}{{d{\Omega
_{DE}}}}\frac{{d{\Omega _{DE}}}}{{dt}} = 2{q^2} + q -
3\frac{{dq}}{{d{\Omega _{DE}}}}\left[ {\frac{{\left( {{\Omega
_{DE}} - {c^2}} \right)\left( {3\left( {1 - {\Omega _{DE}}}
\right) - {b^2}} \right)}}{{1 + {\Omega _{DE}} - 2{c^2}}}}
\right].
\end{equation}

\begin{figure}[htp]\label{fig6}
\centering

\begin{tabular}{cc}

\includegraphics[width=80mm]{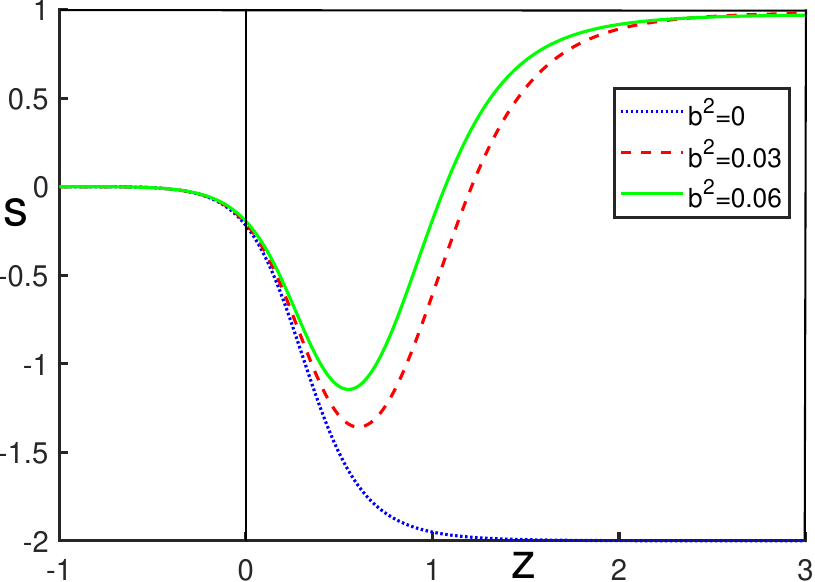}

\includegraphics[width=80mm]{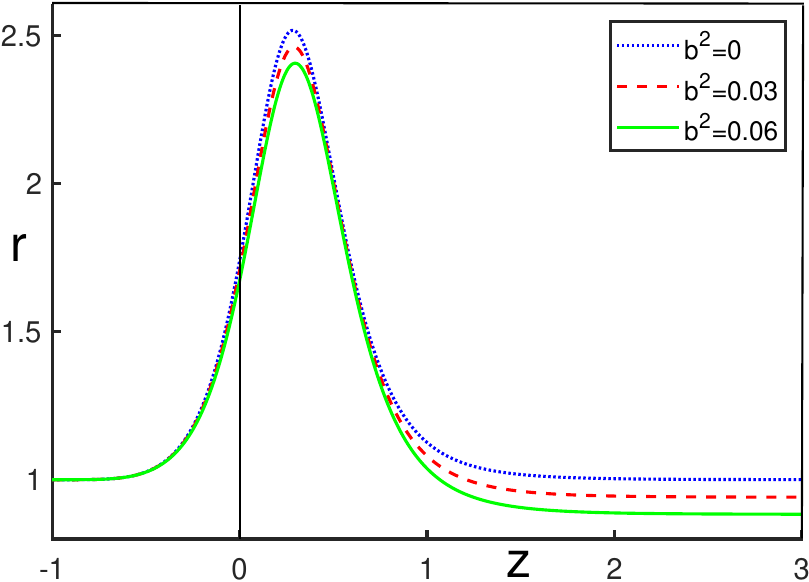} \\

\includegraphics[width=80mm]{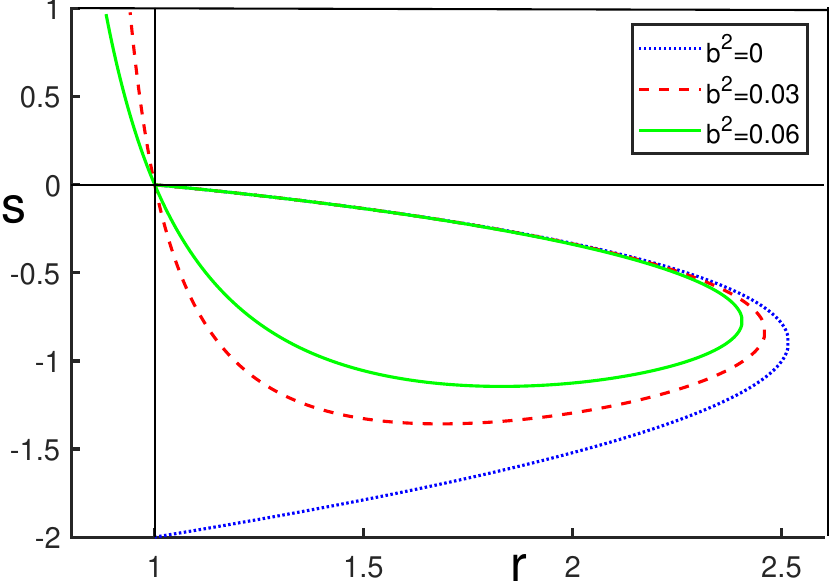}

\includegraphics[width=80mm]{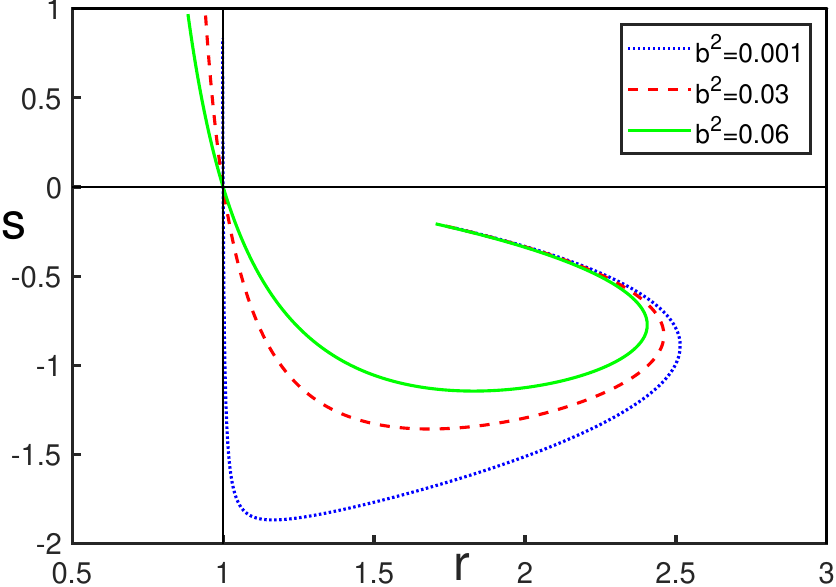}

\end{tabular}
\caption{\scriptsize Evolution of the $s$ (the up and left panel)
and the $r$ (the up and right panel) as a function of redshift,
$z$, and the $s$ (the down and left panel) as a function of , $r$,
for non-interacting and interacting \textit{HDE} for different
choices of ${b^2}$. Here, we set $\Omega_{DE0}$= 0.7, ${c^2} =
0.5$ (${c^2} < 0.65$) and also assume a flat background.}
\end{figure}
Here, we consider the statefinder analysis for KHDE in a flat
background. To this end we plot $r$ and $s$ in terms of $z$ in
Fig.\ref{fig6}. According to upper panels we observe that the
model can catch the point ${1,0}$ in the future(in both
interacting and non-interacting mode irrespective of $b^2$ value)
while at present time the model evidently is distinct from the
$\Lambda$CDM. Looking at $r-z$ and $s-z$ diagram, it can be seen
that behavior of the model is sensitive to the coupling constant,
$b^2$, parameter at past epochs. Lower part of Fig. \ref{fig6}
confirms above description.

\section{Closing remarks} \label{Con}
In most entropy-corrected HDE models, one modifies the energy
density of the HDE and keep the background field equations similar
to standard cosmology. However, using thermodynamics-gravity
conjecture, it has been confirmed that any modification to the
entropy expression, not only change the energy density of HDE
through holographic principle, but also modifies the dynamical
Friedmann equations by applying the first law of thermodynamics on
the apparent horizon of FRW universe.

In this work we have investigated KHDE model with Hubble horizon
as IR cutoff in the background of the modified Kaniadakis
cosmology. We have examined several cases, including DE dominated
universe, a case which the universe is filled with KHDE and DM
which evolves separately as well as an interacting mode. In order
to investigate the cosmological application of KHDE we extracted
the differential equation that determines the evolution of the
effective DE density parameter $\Omega_{DE}$. We first considered
a dark energy (DE) dominated universe and found that this model
mimics the cosmological constant with $w_{DE}=-1$. This implies
that the theoretical origin of the cosmological constant,
$\Lambda$, may be understood through KHDE in Kaniadakis cosmology.
Remarkably, we observe that in the absence of interaction between
DE and dark matter (DM), and in contrast to HDE in standard
cosmology, our model can explain the current acceleration of the
cosmic expansion for Hubble horizon as IR cutoff. We observed
that, regardless of the interaction term between two dark
components, the EoS parameter, $w_{DE}$ and total EoS parameter,
$w_{tot}$, decreases by increasing interaction parameter ($b^2$)
during the history of the universe. $w_{DE}$ crosses the phantom
divide at the present time. We also observed that the $w_{tot}$ of
the KHDE remains in quintessence era, and approaches to the
cosmological constant ($w_{tot}=-1$) at future for non-interacting
while the  in the interacting mode the model $w_{tot 0}<-1$. We
found that decreasing $b^2$ leads to a delay in the cosmic phase
transition during the history of the universe. Also for a fixed
value of Kaniadakis parameter $K$, by increasing $b^2$, the value
of the EoS parameter decreases in all epochs. We have also
explored the squared sound speed stability for this model and
found out that by increasing of $b^2$, the stability of the model
decreases. The statefinder analysis revealed that the KHDE in
interacting and non-interacting mode irrespective of $b^2$ value
catch ${1,0}$ at future epochs while this model behaves well
distinct with respect to $\Lambda$CDM.
\acknowledgments{We are grateful to Shiraz University Research
Council. A. Asvar thanks M. Mohammadi for helpful discussions and
valuable comments.}

\end{document}